
%
%
%
\magnification=1200
\hsize=16.0 truecm
\vsize=22.5 truecm
\baselineskip=18 true pt
\lineskip=2 pt
\footline={\hss\tenrm-- \folio\ --\hss}
\raggedbottom
\def\ltsima{$\; \buildrel < \over \sim \;$}
\def\lsim{\lower.5ex\hbox{\ltsima}}
\def\gtsima{$\; \buildrel > \over \sim \;$}
\def\gsim{\lower.5ex\hbox{\gtsima}}

\def\msole{~M_{\odot}}
\def\nupa{\vfill\eject\noindent}

\def\grbs{$\gamma$-ray bursts}
\def\aa #1 #2  {{A\&A} {#1} #2}
\def\aas #1 #2  {{A\&AS} {#1} #2}
\def\araa #1 #2  {{ARA\&A} {#1} #2}
\def\mon #1 #2  {{MNRAS} {#1} #2}
\def\apj #1 #2  {{ApJ} {#1} #2}
\def\apjs #1 #2  {{ApJS} {#1} #2}
\def\astrj #1 #2  {{AJ} {#1} #2}
\def\nat #1 #2  {{Nat} {#1} #2}
\def\pasj #1 #2  {{PASJ} {#1} #2}
\def\pasp #1 #2  {{PASP} {#1} #2}
\def\mess #1 #2  {{Messenger} {#1} #2 }
\def\ass #1 #2  {{ASS} {#1} #2}
\def\phrevl #1 #2  {{Phys Rev. Lett.} {#1} #2 }
\def\phlB #1 #2  {{Phys. Lett. B} {#1} #2 }
\def\iau #1 #2 {{IAU Circ. No} #1 #2}
\def\europh #1 #2  {{Europhys. News} {#1} #2 }
\def\acta #1 #2  {{Acta Astr.} {#1} #2 }
\def\fundcph #1 #2  {{Fund. Cosmic Phys} {#1} #2 }
\def\ref{\par\noindent\hangindent=.8truecm}

\centerline{\bf MICROLENSING AND HALO MASS IN FORM OF STARS}

\bigskip
\centerline{Sergio Campana$^{\,*}$}

\smallskip
{Osservatorio Astronomico di Brera, Via Brera 28, 20121 Milano, Italy}

\smallskip
{$^*$ affiliated to I.C.R.A.}

\vskip 1truecm

\centerline{\bf Abstract}
\medskip

Paczy\'nski (1986) suggested that ``dark" objects in the halo of our
Galaxy could enhance the luminosity of foreground stars, acting as
gravitational microlenses. Such events has been recently reported by
different collaborations. We assume that these microlensing events are
produced by baryonic objects in the halo of our Galaxy. Rather than
adopting a mean mass for the lensing objects we consider them
distributed along a mass function similar to the disk one. Even if the
number of microlensing events so far detected is not large, we show that
it is possible to constrain the fraction of mass density in the halo of
our Galaxy in form of stars. We estimate that this fraction varies
between 0.1 and 0.8, if halo stars trace the total halo mass density; in
the opposite case this range becomes narrower. If lensing objects have
not been formed apart from the other halo stars, but in a more general
context of star formation, neutron stars must also be present. We
discuss these results in relation to galactic halo models for \grbs.


\vskip 1truecm

{\noindent \bf 1. Introduction}
\bigskip

The nature of dark matter forming the halo of our Galaxy could be
partially resolved by the ongoing experiments of gravitational
microlensing. Paczy\'nski (1986) proposed that compact remnants, such as
black holes and neutron stars, or ``Jupiters" and brown dwarfs
(collectively named as Massive Astronomical Compact Halo Object, MACHO),
could enhance the luminosity of foreground stars by gravitational
microlensing.
Three experiments are searching for these amplification events by
monitoring millions of stars: the American-Australian MACHO
collaboration at Mt. Stromlo Observatory, the French EROS collaboration
at ESO, both monitoring stars in the Large Magellanic Cloud (LMC), and
the American-Polish OGLE collaboration at Las Campa\~nas monitoring the
galactic bulge. The first two collaborations have reported the detection
of one (MACHO; Alcock etal 1993) and two (EROS; Aubourg et al. 1993)
events with the ``canonical" features of microlensing amplification:
symmetry of the light curve around its maximum and achromaticity (i.e.
equal light curve for different colors; Paczy\'nski 1986). A fourth
event has been observed by OGLE collaboration (Udalski et al. 1993), but
this experiment should not directly probe the halo of our Galaxy.
All the lensed LMC stars have almost the same luminosity ($\sim 19.5\
m$) and might also represent a new class of variable stars, mimicking
microlensing events.
Intense monitoring of these stars will prove if we are dealing with a
new class of variable stars or with true microlensing events.

\bigskip
{\noindent \bf 2. The fraction oh halo mass in form of stars}
\medskip

Turner (1993), assuming a characteristic mass for the lensing object of
$0.1\msole$, derived that the fraction of mass in the halo in form of
stars, $f$, must be between $0.1-0.5$, in order to be consistent with
the observed number of microlensing events. This value of $f$ should
provide a good approximation: from the duration of the events one can
derive the most probable mass of the lensing object, which turns out to
be a broad gaussian between $0.01-1\msole$, peaked at $\sim 0.1\msole$
(Griest 1991). Detailed calculations on the mass of the lensing objects
give $0.12^{+0.26}_{-0.08}\msole$ for the MACHO event,
$0.31^{+0.64}_{-0.20} \msole$ and $0.38^{+0.82}_{-0.25}\msole$ for the
EROS events; the average mass is estimated in $0.14\msole$ (Jetzer \&
Mass\`o 1994).

Here we calculate the fraction of halo mass in stars to be consistent
with the observed number of microlensing events, taking into account a
distribution of masses, rather than an unique value. We would have to
use the halo mass function but it is unknown, therefore we adopt the
disk Initial Mass Function (IMF) as a good approximation.
(It can be shown that high mass stars ($M\gsim 1\msole$) practically do not
contribute to the number of lensing events observed with current
experiments, which is mainly determined by low mass objects).
The disk IMF for masses greater than about one solar mass is known with
good accuracy, problems arise for lower masses. Observational results on
the stellar birthrate in the solar neighborhood require the presence of
a ``knee" in order to decrease the number of low mass stars (Miller \&
Scalo 1979). The most recent work (Tinney 1993) indicates the inflection
point at $\sim 0.25\msole$; below this value the IMF could be either
slowly rising or flat (Pound \& Blitz 1993; Tinney 1993). We adopt the
IMF proposed by Ferrini and co-workers (1990),
which is slightly steeper than the classical Salpeter's at high masses;
at the low mass end shows a ``knee", at about $0.3\msole$, and is slowly
rising below. We take two different low mass ends for the IMF, namely
$M_L=0.1\msole$ and $M_L=0.001\msole$, in order to underline the
importance of small mass objects to produce microlensing events. (We
have also considered $M_L=10^{-4}\msole$, but results are very similar
to the case of $M_L=0.001\msole$.) A flat IMF below $0.25\msole$
provides $\sim 15\%$ and $\sim 50\%$ less microlensing events than the
adopted one, for $M_L=0.1\msole$ and $M_L=0.001\msole$ respectively.

The number of microlensing events depends also on the spatial
distribution of lensing objects observed from the Sun. We do not know if
stars in the halo trace the total halo mass density. We consider two
contributions to the total density, one deriving from halo stars and the
other from matter unable to produce microlensing events.
For both these distributions we assume an isothermal profile
characterized by a core radius: $r_c$ for halo stars and $R_c$ for
non-lensing halo matter. The total halo density can be expressed as:

$$\rho(r)=\rho_0\, \Bigl[ \, f\,{{r_0^2+r_c^2}\over {r^2+r_c^2}} +
(1-f)\,{{r_0^2+R_c^2}\over {r^2+R_c^2}} \, \Bigr] \eqno(1)$$

\noindent where $\rho_0=7.9\times10^{-3}\msole\,{\rm pc^{-3}}$ is the
local dark matter density (Flores 1988), $r_0=8.5$~kpc is the distance
between the sun and the galactic center, and $r$ measures distances from
the Sun (LMC distance is $r_{LMC}=50$~kpc).
We consider a halo extending, at least, up to the LMC: indications of
large halo extensions come from the observational requirement of metal
ejection from protogalaxies into the intergalactic medium (Hattori \&
Terasawa 1993) and by the interpretation of the Magellanic stream
(Binney \& Tremaine 1986). By assuming equation (1) we neglect a
flattening of the halo at large distances (Sackett \& Gould 1993) and
contributions from a halo around the LMC (Gould 1993).

The first term in equation (1) refers to stars in the halo which are
responsible for microlensing events, the other one takes into account
the presence of matter which is not responsible for microlensing events
and could be either baryonic or non baryonic. The halo core radius has
been estimated in $2-8$ kpc (Caldwell \& Ostriker 1981; Bachall, Schmidt
\& Soneira 1983), therefore for $f\ll 1$ the non-lensing matter is
dominant and we take $R_c\sim 5$\ kpc (and leave $r_c$ free); for $f\sim
1$ or if halo stars trace the total halo density we must require
$r_c\sim 5$\ kpc.

Following Griest (1991) and taking into consideration the number of LMC
stars monitored, the percentage of data effectively analysed, the time
efficiency of the observations (see Table$\,$1),
we obtain the predicted number of events
relative to MACHO and EROS experiments, depending only on the the
fraction of the halo mass in stars and the core radius of the star
distribution.
By deriving fiducial intervals for the microlensing events by means of
the Poisson statistics (90\%), we report in Tables 2 and 3 the limits on
the fraction of the halo mass in form of stars. These values refer to a
stationary star and observer. Including a velocity distribution of halo
objects and non-zero velocity of the Sun, star and Earth, results change
very little (Griest 1991). We estimate also the most probable mass of
the lensing stars in $0.13\msole$.

In Table 2 we consider halo stars tracing the total distribution of halo
mass. We obtain a large range for the fraction of halo mass in form of
stars $0.1\lsim f \lsim 0.8$, consistent with the one found by Turner
(1993) using simplified arguments. In Table 3 we consider a contribution
to the halo from non-lensing mass and leave open the possibility that
halo stars do not trace the total distribution (i.e. $f\ll 1$). In this
case constraints of $f$ are more stringent due to the lower mass
available for producing microlensing events. We take $R_c=5$\ kpc and
derive limits on the fraction of halo mass in form of stars as a
function of the star core radius $r_c$. We can allow much larger star
core radii, but for small values of $r_c$ stars trace the total halo
mass distribution and we obtain high values of $f$ (contrary to our
assumption), instead for large star core radii we have $f\sim 0.3-0.4$.

\bigskip
{\bf 3. CONCLUSIONS}
\medskip

The upper limit on $f$ indicates that the mass of our halo could be
almost completely baryonic in form; at the same time the lower limit
indicates that there must be some baryonic matter in form of stars to
produce the observed microlensing events.
If this mass has been repartee into stars by means of an ``usual" IMF,
we have that the most probable objects that have been formed are low
mass stars, while high mass objects are rare. Therefore, it is not
surprising that the mean lens mass is about $0.1\msole$, moreover being
microlensing experiments biased towards low masses because of the
temporal coverage (Griest 1991). Following this line of reasoning we
have, as a by product of the halo stellar evolution, a number of compact
remnants which are too few to give microlensing effects in respect to
low mass stars, but which should be responsible of \grbs.

BATSE results (Meegan et al. 1993), imply that the neutron star core
radius is greater than $\sim 20$\ kpc in order to have an isotropic
distribution of \grbs\ (Mao \& Paczy\'nski 1992; Brainerd 1992; Hakkila
et al. 1994).
Considering the
observed number of microlensing events we have to face two different
possibilities: stars in the halo of our Galaxy may or may not trace the
total halo density. In the first case we can not allow large star core
radii because of the total halo core radius estimate
(Caldwell \& Ostriker 1981; Bachall et al. 1983)
and therefore \grbs\ are likely cosmological. We
expect $0.1\lsim f \lsim 0.8$. In the opposite case we must require
large halo star core radii in order to have an isotropic distribution of
\grbs\ and we have $f\sim 0.3-0.4$. New microlensing events, as well as
a larger sample of \grbs, will further constrain the value of $f$,
providing indications on the nature of matter in the halo of our Galaxy
and on the origin of \grbs.

\vskip 1.5truecm
\noindent ACKNOWLEDGMENTS
\bigskip
I thank L. Stella for useful discussions and suggestions.
Support for this work was partially provided by SISSA/ISAS, I thank A.
Treves for this opportunity.

\nupa
\vskip 2truecm
\centerline{\bf REFERENCES}
\vskip 1truecm

\ref
Alcock, C. et al., 1993, \nat 365 621

\ref
Aubourg, E. et al., 1992, \mess 72 20

\ref
Aubourg, E. et al., 1993, \nat 365 623

\ref
Bahcall, J.N., Schmidt, M. \& Soneira, R.M., 1983, \apj 265 730

\ref
Binney, J. \& Tremaine, S., 1986, {\it Galactic Dynamics} (Princeton
University Press)

\ref
Brainerd, J.J., 1992, \nat 355 522

\ref
Caldwell, J.A.R. \& Ostriker, J.P., 1981, \apj 251 61

\ref
Ferrini, F., Matteucci, F., Pardi, M.C. \& Penco, U., 1992, \apj 387 138

\ref
Flores, R.A., 1988, \phlB 215 73

\ref
Gould, A., 1993, \apj 404 451

\ref
Griest, K., 1991, \apj 366 412

\ref
Hakkila, J. et al., 1994, \apj 422 659

\ref
Hattori, M. \& Terasawa, N., 1993, \apj 406 L55

\ref
Jetzer, Ph. \& Muss\'o, E., 1994, {\it Phys. Lett. B}, in press

\ref
Mao, S. \& Paczy\'nski, B., 1992, \apj 389 L13

\ref
Meegan, C.A. et al. in AIP Conf. Proc., 1993, {\it Compton Gamma-Ray
Observatory} (eds Friedlander, M., Gehrels, N. \& Macomb, D.), in
press

\ref
Miller, G.E. \& Scalo, J.M., 1979, \apjs 41 513

\ref
Paczy\'nski, B., 1986, \apj 304 1

\ref
Pound, M.W. \& Blitz, L., 1993, \apj 418 328

\ref
Sackett, P.D. \& Gould, A., 1993, \apj 419 648

\ref
Tinney, C.G., 1993, \apj 414 279

\ref
Turner, M.S., 1993, astro-ph/9310019

\ref
Udalski, A. et al., 1993, \acta 43 289

\nupa
%
\message{S-Tables Macro v1.0, ACS, TAMU (RANHELP@VENUS.TAMU.EDU)}
%
%
\newhelp\stablestylehelp{You must choose a style between 0 and 3.}%
\newhelp\stablelinehelp{You should not use special hrules when stretching
a table.}%
\newhelp\stablesmultiplehelp{You have tried to place an S-Table inside another
S-Table.  I would recommend not going on.}%
%
%
\newdimen\stablesthinline
\stablesthinline=0.4pt
\newdimen\stablesthickline
\stablesthickline=1pt
%
%
\newif\ifstablesborderthin
\stablesborderthinfalse
\newif\ifstablesinternalthin
\stablesinternalthintrue
\newif\ifstablesomit
\newif\ifstablemode
\newif\ifstablesright
\stablesrightfalse
%
%
\newdimen\stablesbaselineskip
\newdimen\stableslineskip
\newdimen\stableslineskiplimit
%
%
\newcount\stablesmode
\newcount\stableslines
\newcount\stablestemp
\stablestemp=3
\newcount\stablescount
\stablescount=0
\newcount\stableslinet
\stableslinet=0
%
%
%
\newcount\stablestyle
\stablestyle=0
%
%
\def\stablesleft{\quad\hfil}%
\def\stablesright{\hfil\quad}%
%
%
\catcode`\|=\active%
%
%
\newcount\stablestrutsize
\newbox\stablestrutbox
\setbox\stablestrutbox=\hbox{\vrule height10pt depth5pt width0pt}
\def\stablestrut{\relax\ifmmode%
                         \copy\stablestrutbox%
                       \else%
                         \unhcopy\stablestrutbox%
                       \fi}%
%
%
\newdimen\stablesborderwidth
\newdimen\stablesinternalwidth
\newdimen\stablesdummy
\newcount\stablesdummyc
\newif\ifstablesin
\stablesinfalse
%
%
\def\begintable{\stablestart%
  \stablemodetrue%
  \stablesadj%
  \halign%
  \stablesdef}%
\def\stablesadj{%
  \ifcase\stablestyle%
    \hbox to \hsize\bgroup\hss\vbox\bgroup%
  \or%
    \hbox to \hsize\bgroup\vbox\bgroup%
  \or%
    \hbox to \hsize\bgroup\hss\vbox\bgroup%
  \or%
    \hbox\bgroup\vbox\bgroup%
  \else%
    \errhelp=\stablestylehelp%
    \errmessage{Invalid style selected, using default}%
    \hbox to \hsize\bgroup\hss\vbox\bgroup%
  \fi}%
\def\stablesend{\egroup%
  \ifcase\stablestyle%
    \hss\egroup%
  \or%
    \hss\egroup%
  \or%
    \egroup%
  \or%
    \egroup%
  \else%
    \hss\egroup%
  \fi}%
\def\stablestart{%
  \ifstablesin%
    \errhelp=\stablesmultiplehelp%
    \errmessage{An S-Table cannot be placed within an S-Table!}%
  \fi
  \global\stablesintrue%
  \global\advance\stablescount by 1%
  \message{<S-Tables Generating Table \number\stablescount}%
  \begingroup%
  \stablestrutsize=\ht\stablestrutbox%
  \advance\stablestrutsize by \dp\stablestrutbox%
  \ifstablesborderthin%
    \stablesborderwidth=\stablesthinline%
  \else%
    \stablesborderwidth=\stablesthickline%
  \fi%
  \ifstablesinternalthin%
    \stablesinternalwidth=\stablesthinline%
  \else%
    \stablesinternalwidth=\stablesthickline%
  \fi%
  \tabskip=0pt%
  \stablesbaselineskip=\baselineskip%
  \stableslineskip=\lineskip%
  \stableslineskiplimit=\lineskiplimit%
  \offinterlineskip%
  \def\borderrule{\vrule width \stablesborderwidth}%
  \def\internalrule{\vrule width \stablesinternalwidth}%
  \def\thinline{\noalign{\hrule height \stablesthinline}}%
  \def\thickline{\noalign{\hrule height \stablesthickline}}%
  \def\trule{\omit\leaders\hrule height \stablesthinline\hfill}%
  \def\ttrule{\omit\leaders\hrule height \stablesthickline\hfill}%
  \def\tttrule##1{\omit\leaders\hrule height ##1\hfill}%
  \def\stablesel{&\omit\global\stablesmode=0%
    \global\advance\stableslines by 1\borderrule\hfil\cr}%
  \def\el{\stablesel&}%
  \def\elt{\stablesel\thinline&}%
  \def\eltt{\stablesel\thickline&}%
  \def\elttt##1{\stablesel\noalign{\hrule height ##1}&}%
  \def\elspec{&\omit\hfil\borderrule\cr\omit\borderrule&%
              \ifstablemode%
              \else%
                \errhelp=\stablelinehelp%
                \errmessage{Special ruling will not display properly}%
              \fi}%
  \def\stmultispan##1{\mscount=##1 \loop\ifnum\mscount>3 \stspan\repeat}%
  \def\stspan{\span\omit \advance\mscount by -1}%
  \def\multicolumn##1{\omit\multiply\stablestemp by ##1%
     \stmultispan{\stablestemp}%
     \advance\stablesmode by ##1%
     \advance\stablesmode by -1%
     \stablestemp=3}%
  \def\multirow##1{\stablesdummyc=##1\parindent=0pt\setbox0\hbox\bgroup%
    \aftergroup\emultirow\let\temp=}
  \def\emultirow{\setbox1\vbox to\stablesdummyc\stablestrutsize%
    {\hsize\wd0\vfil\box0\vfil}%
    \ht1=\ht\stablestrutbox%
    \dp1=\dp\stablestrutbox%
    \box1}%
  \def\stpar##1{\vtop\bgroup\hsize ##1%
     \baselineskip=\stablesbaselineskip%
     \lineskip=\stableslineskip%
     \lineskiplimit=\stableslineskiplimit\bgroup\aftergroup\estpar\let\temp=}%
  \def\estpar{\vskip 6pt\egroup}%
  \def\stparrow##1##2{\stablesdummy=##2%
     \setbox0=\vtop to ##1\stablestrutsize\bgroup%
     \hsize\stablesdummy%
     \baselineskip=\stablesbaselineskip%
     \lineskip=\stableslineskip%
     \lineskiplimit=\stableslineskiplimit%
     \bgroup\vfil\aftergroup\estparrow%
     \let\temp=}%
  \def\estparrow{\vfil\egroup%
     \ht0=\ht\stablestrutbox%
     \dp0=\dp\stablestrutbox%
     \wd0=\stablesdummy%
     \box0}%
  \def|{\global\advance\stablesmode by 1&&&}%
  \def\|{\global\advance\stablesmode by 1&\omit\vrule width 0pt%
         \hfil&&}%
  \def\vt{\global\advance\stablesmode by 1&\omit\vrule width \stablesthinline%
          \hfil&&}%
  \def\vtt{\global\advance\stablesmode by 1&\omit\vrule width
\stablesthickline%
          \hfil&&}%
  \def\vttt##1{\global\advance\stablesmode by 1&\omit\vrule width ##1%
          \hfil&&}%
  \def\vtr{\global\advance\stablesmode by 1&\omit\hfil\vrule width%
           \stablesthinline&&}%
  \def\vttr{\global\advance\stablesmode by 1&\omit\hfil\vrule width%
            \stablesthickline&&}%
  \def\vtttr##1{\global\advance\stablesmode by 1&\omit\hfil\vrule width ##1&&}%
  \stableslines=0%
  \stablesomitfalse}
\def\stablesdef{\bgroup\stablestrut\borderrule##\tabskip=0pt plus 1fil%
  &\stablesleft##\stablesright%
  &##\ifstablesright\hfill\fi\internalrule\ifstablesright\else\hfill\fi%
  \tabskip 0pt&&##\hfil\tabskip=0pt plus 1fil%
  &\stablesleft##\stablesright%
  &##\ifstablesright\hfill\fi\internalrule\ifstablesright\else\hfill\fi%
  \tabskip=0pt\cr%
  \ifstablesborderthin%
    \thinline%
  \else%
    \thickline%
  \fi&%
}%
\def\endtable{\advance\stableslines by 1\advance\stablesmode by 1%
   \message{- Rows: \number\stableslines, Columns:  \number\stablesmode>}%
   \stablesel%
   \ifstablesborderthin%
     \thinline%
   \else%
     \thickline%
   \fi%
   \egroup\stablesend%
\endgroup%
\global\stablesinfalse}
%
%
\hsize= 7truein
\vsize= 8truein
{\centerline{{\bf TABLE 1}: MACHO and EROS parameters}}
\vskip 0.5truecm
\begintable
Parameters| MACHO$^1$ |EROS$^{2,\,3}$\eltt
Time of observations | 1 yr  | 1.44 yr \elt
Minimum time between observations (mean)| 1.5 d | 1.7 d\elt
Number of monitored stars | $1.8\times 10^6$ | $8 \times 10^6$\elt
Efficiency (binary, variable, | 0.5 | 0.25\el
sufficiently luminous stars) ||\elt
Percentage of analyzed observations | 0.15 | 0.4\endtable
\bigskip

\noindent {1 Alcock et al. 1993}

\noindent {2 Aubourg et al. 1992}

\noindent {3 Aubourg et al. 1993}

\vskip 4truecm

\nupa
\hsize= 7truein
\vsize= 8truein
\nopagenumbers
{\centerline{{\bf TABLE 2}}}
\vskip 0.5truecm
\begintable
Star Core \vtt MACHO | EROS \vtt MACHO | EROS \el
radius ($r_c$) \vtt $M_L=0.1\msole$|$M_L=0.1\msole$ \vtt $M_L=0.001\msole$|
$M_L=0.001\msole$\eltt
{}~\vtt~|~\vtt~|~\el
2 kpc \vtt 0.32 -- 1 | 0.20 -- 1    \vtt
           0.19 -- 1 | 0.12 -- 0.93 \elt
5 kpc \vtt 0.29 -- 1 | 0.18 -- 1    \vtt
           0.17 -- 1 | 0.11 -- 0.82 \elt
10 kpc \vtt 0.22 -- 1 | 0.14 -- 1    \vtt
            0.13 -- 1 | 0.08 -- 0.63 \endtable

\bigskip
\noindent {Limits (90\%) on the fraction of halo mass in form of stars
consistent with the observed number of microlensing events, for a star
distribution which traces the total halo mass.

\vskip 3truecm

{\centerline{{\bf TABLE 3}}}
\vskip 0.5truecm
\begintable
Star Core \vtt MACHO | EROS \vtt MACHO | EROS \el
radius ($r_c$) \vtt $M_L=0.1\msole$|$M_L=0.1\msole$ \vtt $M_L=0.001\msole$|
$M_L=0.001\msole$\eltt
{}~\vtt~|~\vtt~|~\el
2 kpc \vtt 0.68 -- 1 | 0.62 -- 1    \vtt
           0.62 -- 1 | 0.59 -- 0.97 \elt
5 kpc \vtt 0.64 -- 1 | 0.59 -- 1    \vtt
           0.58 -- 1 | 0.55 -- 0.91 \elt
10 kpc \vtt 0.56 -- 1 | 0.51 -- 1    \vtt
            0.51 -- 1 | 0.48 -- 0.79 \elt
20 kpc \vtt 0.43 -- 1 | 0.40 -- 0.81 \vtt
            0.39 -- 1 | 0.37 -- 0.61 \elt
50 kpc \vtt 0.32 -- 1 | 0.29 -- 0.59 \vtt
            0.29 -- 1 | 0.27 -- 0.45 \endtable

\bigskip
\noindent {Limits (90\%) on the fraction of halo mass in form of stars
which do not trace the total halo mass distribution.
The core radius of the non-lensing matter distribution has been fixed
to $R_c=5$\ kpc.

\end